\documentclass[aps,prl,twocolumn,showpacs,superscriptaddress,groupedaddress]{revtex4}  % for review and submission
\usepackage{graphicx}  % needed for figures
\usepackage{dcolumn}   % needed for some tables
\usepackage{bm}        % for math
\usepackage{amssymb}   % for math

\begin{document}

% The following information is for internal review, please remove them for submission
\widetext
%\leftline{Version xx as of \today}
%\leftline{Primary authors: Joe E. Physics}
%\leftline{To be submitted to (PRL, PRD-RC, PRD, PLB; choose one.)}
%\leftline{Comment to {\tt d0-run2eb-nnn@fnal.gov} by xxx, yyy}
%\centerline{\em D\O\ INTERNAL DOCUMENT -- NOT FOR PUBLIC DISTRIBUTION}
\centerline{\em Under consideration for publication in XXX}
% the following line is for submission, including submission to the arXiv!!
%\hspace{5.2in} \mbox{Fermilab-Pub-04/xxx-E}

\title{Symmetry and conservation principles of evolution of general Rayleigh-Taylor mixing fronts}% Force line breaks with \\
%\thanks{Arguments in favor of understanding the universality of turbulent boundary layer}
%\input author_list.tex       % D0 authors (remove the first 3 lines
                             % of this file prior to submission, they
                             % contain a time stamp for the authorlist)
                             % (includes institutions and visitors)

\author{You-sheng Zhang}
\author{Zhi-wei He}
\author{Fu-jie Gao}
\affiliation{%
Institute of applied physics and computational mathematics, Beijing 100094, China
}%

%\author{Fazle Hussain}
%\affiliation{Department of Mechanical Engineering, University of Houston, Houston, TX 77204-4006, %USA}%

\author{Xin-liang Li}
  \affiliation{Institute of Mechanics, Chinese Academy of Sciences, Beijing 100190, China.}%

\author{Bao-lin Tian}
\email{tian_baolin@iapcm.ac.cn}
\affiliation{%
Institute of applied physics and computational mathematics, Beijing 100094, China
}%

\date{\today}

\begin{abstract}
%\normalsize
\small
A theory determining the evolution of general Rayleigh-Taylor mixing fronts is established to reproduce firstly all of the documented experiments conducted for diverse acceleration histories and all density ratios. The theory is established in terms of the fundamental conservation and symmetry principles, with special consideration given to the symmetry breaking of the density fields occurring in actual flows. The results reveal the sensitivity/insensitivity of the evolution of a mixing front neighbouring light/heavy fluid to the degree of symmetry breaking, and also explain the distinct evolutions in two experiments with the same configurations.
\end{abstract}

%Use showkeys class option if keyword
\maketitle
%\tableofcontents

\begin{figure}[t]
\centering
\includegraphics[width=8.5cm]{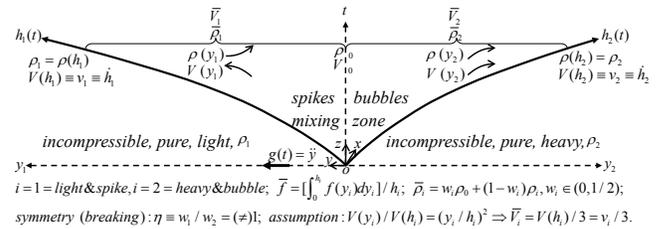}
\caption{ \label{fig1} Problem set up, notations, ideas, and assumption for the current theory.}
\end{figure}

As shown in Fig.\ref{fig1}, when two fluids of density $\rho_i$ (i = 1 = light, i = 2 = heavy) are separated by a perturbed interface and are accelerated in the direction opposite to that of the density gradient, Rayleigh-Taylor (RT) instability occurs and develops rapidly into turbulent mixing consisting with bubbles/spikes mixing zone (BMZ/SMZ) \cite{Cheng2002}. The mixing occurs ubiquitously in systems extending from the micro to astrophysical scales \cite{Cabot2006}. As the simplest and primary descriptor of the mixing process, the evolution of the two edges of the mixing zone (i.e. mixing fronts $h_i(t)$, i = 1 = spikes, i = 2 = bubbles) plays a notable role \cite{Cheng2002} in many natural phenomena (e.g., supernova explosions \cite{Burrows2000}) and engineering applications (e.g., inertial confinement fusion \cite{Petrasso1994}). The scenarios generally involve complex varying acceleration histories $g(t)$ and widely varying density ratios $R \equiv {\rho _2}/{\rho _1}$, two dominant factors \cite{Dimonte2000} affecting $h_i(t)$.

\begin{figure}[t]
\centering
\includegraphics[width=8.5cm,trim=0.15cm 0 0 1.00cm,clip]{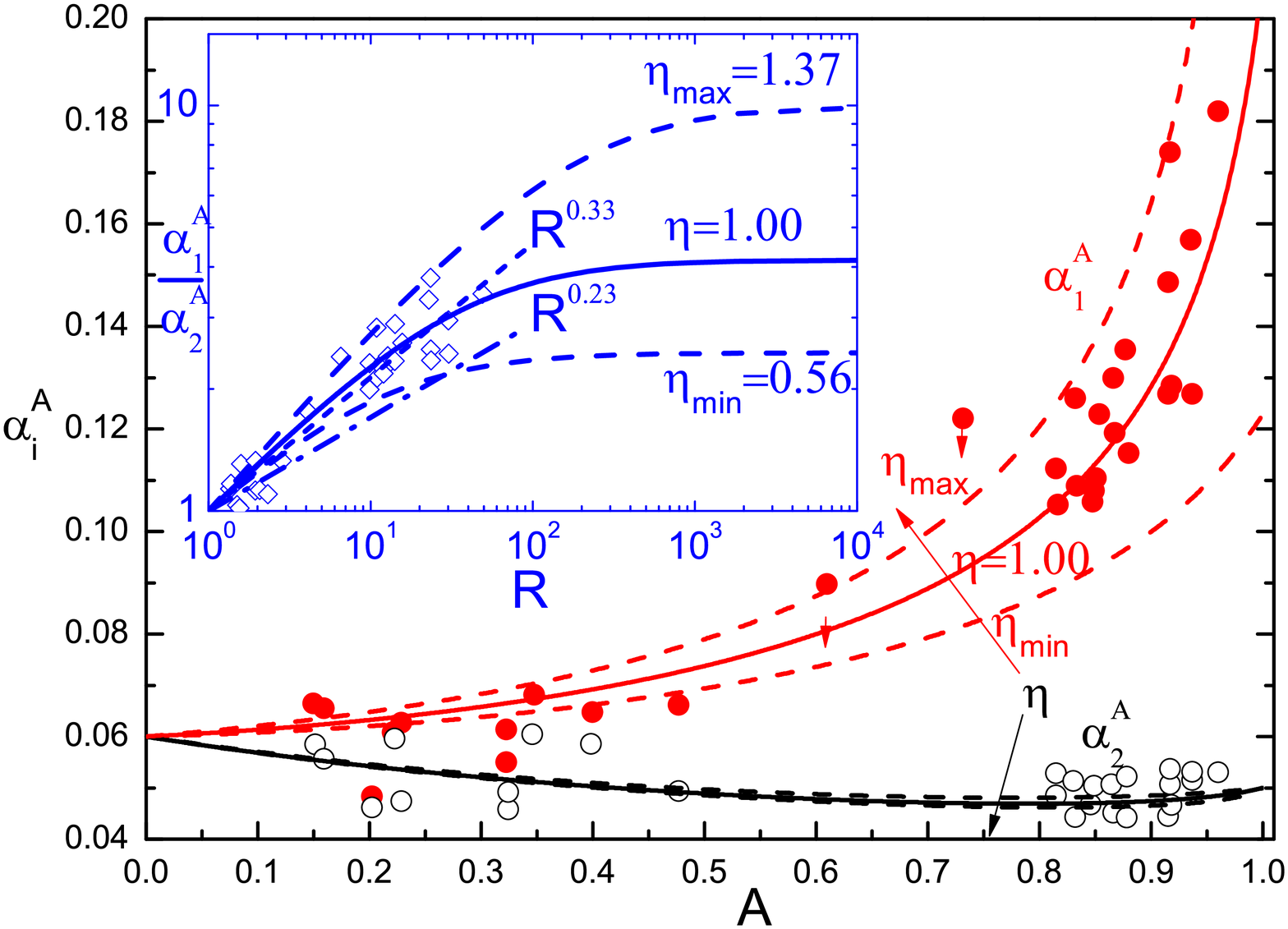}
\caption{ \label{fig2} (color online). The comparison of $\alpha_1^A$ (red),$\alpha_2^A$ (black) and $\alpha_1^A/\alpha_2^A$ (inset) between the experiments (symbols \cite{Dimonte2000exp}) and the theoretical predictions (lines) for problems with constant $g$ and at all $R$. The symbols with $\downarrow$ denote the two over-measured data \cite{Dimonte2000exp}. In inset, the short dashed line and dash-dot line show the empirical formula given by Youngs' \cite{Youngs1989,Youngs2013} and Dimonte \cite{Dimonte2000exp}, respectively.}
\end{figure}

To predict the $h_i(t)$ of general RT problem, many models have been developed in the last few decades, including the bubble-competition model \cite{Alon1995}, energy-transfer model \cite{Ramshaw1998}, stationary-centroid model \cite{Glimm1998} and buoyancy-drag models \cite{Youngs1990,Dimonte2000exp,Dimonte2000,Cheng2000,Cheng2002}. However, no model has completely \cite{Dimonte2000} reproduced the observed $h_i(t)$ \cite{Youngs1989,Dimonte1996,Dimonte2000exp,Dimonte2007}. In fact, even for the simplest RT problem with constant $g$, previous models do not predict satisfactorily. For example, the models produced only one pair of quadratic-growth-coefficients $\alpha _i^A \equiv {h_i}/(Ag{t^2})$, while few pairs of $\alpha _i^A$ were observed \cite{Dimonte2000exp} (see Fig.\ref{fig2}), where $A \equiv (R - 1)/(R + 1)$ is the Atwood number.

In this letter, we describe a new theory for the general incompressible RT problem with the fundamental conservation and symmetry principles. The theory is validated by the series of experiments \cite{Dimonte1996,Dimonte2000exp,Dimonte2007}. Furthermore, it reveals that the $h_{i}(t)$ may be affected by initial perturbations and fluid properties, but governed essentially by mass, momentum conservation and Newton's second law.

We describe our theory with Fig.1. In Fig.1, the $o-xyz$ is a non-inertial reference frame fixed at the initially unperturbed interface\cite{Dimonte2000exp}, denoted by subscript $0$. The $g(t)$ is directed along the $y$ axis, i.e. $\ddot y = g(t)$. Mixing is assumed to be statistically homogeneous in the $x$ and $z$ directions \cite{Glimm1998,Dimonte2000}, with the cross-sectional area set to unity. Consequently, quantities depend only on $y$ and $t$. The $h_i$ quantifies the distance of the interface to the mixing fronts, defined by the locations with concentration $c=1\%(99\%)$ \cite{Dimonte2004}. $V(y)$ quantifies the propagation speed of an iso-concentration surface \cite{Glimm1998} at $y$, with $ V({h_i})\equiv {v_i} \equiv {\dot h_i}$.  Since the density profile $\rho (y)$ in RT problem increases monotonically from $\rho_1$, to $\rho_0$, and to $\rho_2$, one thus has ${\bar \rho _i} = {w_i}{\rho _0} + (1 - {w_i}){\rho _i}$, where $\bar f \equiv [\int_0^{{h_i}} {f({y_i})} d{y_i}]/{h_i}$ defines a volume-average and $w_i$ is called a mixing weight. Since $\rho ({y_i})$ transitions smoothly near $h_i$, $w_i$ should be less than 1/2, i.e. the mixing weight of the linearly varying density profile.

We first introduce the concept of symmetry (breaking) of the density fields as follows: density fields in BMZ and SMZ are said to be in symmetry (breaking) if symmetry breaking factor $\eta  \equiv {w_1}/{w_2} =(\neq)1$. Based on this concept, we establish our theory with conservation principles. First, the conservation of mass \cite{Dimonte2000} requires
\begin{eqnarray}
{\rho _1}{h_1} + {\rho _2}{h_2} = {\bar \rho _1}{h_1} + {\bar \rho _2}{h_2},
\label{eq1}
\end{eqnarray}
Second, given the success of the momentum-driven viewpoint \cite{Sreenivasan2013} in understanding RT mixing and that of the stationary centroid hypothesis in predicting the $h_i(t)$ of constant acceleration RT problems \cite{Glimm1998,Cheng1999,Dimonte2000}, it seem plausible for an approximation of the vanishing resultant force on the entire mixing region, resulting in the quasi-conservation of momentum (similar to the stationary centroid hypothesis):
\begin{eqnarray}
{\bar {\rho}{_1}}{h{_1}}{\dot {\bar V{}}_1} = {\bar {\rho}{_2}}{h{_2}}{\dot {\bar V{}}_2},
\label{eq2}
\end{eqnarray}
which is established by regarding the BMZ/SMZ as a particle. Eq.(\ref{eq2}) implies that the evolutions of $h_1(t)$ and $h_2(t)$ depend on each other, and thus should not be predicted with the two independent equations of the previous models \cite{Youngs1990,Cheng2000,Dimonte2000,Dimonte2000exp,Cheng2002}. Consequently, one additional evolution equation is needed. Given that the bubble structure is independent of $R$ \cite{Alon1995}, we prefer to establish an evolution equation for $h_2(t)$ to avoid  $R$-dependent parameters. For BMZ, it incurs three forces, namely, a buoyancy force ${f_b} = {\rho _2}{h_2}g(t)$, an inertial force ${f_i} = \bar \rho {_2}{h_2}g(t)$, and a drag force ${f_d} = {C_d}{\rho _2}{v_2}|{v_2}|$ \cite{Dimonte2000}, where $C_d$ is the drag coefficient. Applying Newton's second law to BMZ gives
\begin{eqnarray}
{\bar \rho _2}{h_2}{\dot {\bar V}_2} = {f_b} - {f_i} - {f_d}.
\label{eq3}
\end{eqnarray}

In Eq.(\ref{eq2})-(\ref{eq3}), we use the volume-averaged ${\bar V_i}$ to quantify the rate of change of the momentum, instead of using the local $v_i$ in previous models\cite{Youngs1990,Cheng2000,Dimonte2000,Dimonte2000exp,Cheng2002}. In physics, this is more reasonable since the entire BMZ is regarded as being a particle, such that ${\bar V_i}$ should be used. Due to this, however, Eq.(\ref{eq1})-(\ref{eq3}) become unclosed, such that an assumption, with which the relationship between ${\bar V_i}$ and $v_i$ can be derived, is needed. We notice that a reasonable assumption should meet the two following physical intuitions: (1) for unity  $R$, due to symmetry, $V(y_i)$ should increase monotonically from 0 at $y=0$  to $V(h_i)$  at $y=h_i$ ; (2) for any value of  $R$, due to continuity,  ${{(dV/dy)}|_{y \to {0^ + }}}$ $= {{(dV/dy)}|_{y \to {0^ - }}}$. Therefore, the simplest assumption is $V({y_i})/V({h_i})$ $= {({y_i}/{h_i})^2}$, giving ${\bar V_i} = {v_i}/3$  and the final evolution equation:
\begin{eqnarray}
\gamma \chi {\dot v_1} = {\dot v_2} = \beta Ag(t) - C\phi {v_2}|{v_2}|/{h_2},
\label{eq4}
\end{eqnarray}
where $\chi\equiv{h_1}/{h_2}$, $C\equiv3{C_d}$, $\phi\equiv{\rho _2}/{\bar \rho _2} = R(1{\rm{ + }}\chi \eta )/\Theta$, $\beta\equiv3(\phi  - 1)/A = 3{w_2}\chi \eta (R + 1)/\Theta$, $\gamma\equiv {\bar \rho _1}/{\bar \rho _2} = [(R - 1){w_2}\eta  + (1{\rm{ + }}\chi \eta )]/\Theta$, and $\Theta  \equiv R(1 + \chi \eta ) + {w_2}\chi \eta (1 - R)$. Three parameters,$\eta ,{w_2},C$, are incorporated into the current theory. Due to the abovementioned $R$-independent bubble structure, the parameters are postulated to be $R$-independent and are determined with four steps.

Step I: Obtain the exact quadratic solution of Eq.(\ref{eq4}) for problem with constant $g=g_0$, initial values of ${h_{i0}} = 0$ and  ${v_{i0}} = 0$ to give (see Ref.\cite{Dimonte2000} for more information)
\begin{eqnarray}
{h_i} = \alpha _i^AAg{t^2},\alpha _2^A = \beta /(2 + 4{C_{{g_0}}}\phi ),\alpha _1^A = \chi _{{g_0}}^A\alpha _2^A,
\label{eq5}
\end{eqnarray}
where  $\chi _{{g_0}}^A$, equivalent to $\alpha _1^A/\alpha _2^A$ or ${h_1}/{h_2}$, are determined as follows. Substituting the quadratic solution to the first equality of Eq. (4) yields $\gamma {(\chi _{{g_0}}^A)^2} = 1$ and then $\sum_{m=0}^{3} {{a_m}{{(\chi _{{g_0}}^A)}^m}}  = 0$, for which the positive real root gives $\chi _{{g_0}}^A $ $= [B_1^{1/2}(\cos \theta  + \sqrt 3 \sin \theta )$ $- {a_2}]/(3{a_3})$ with ${a_0} =  - R$, ${a_1} = {w_2}\eta (R - 1) - R\eta$, ${a_2} = (R - 1){w_2}\eta  + 1$, ${a_3} = \eta$, ${B_1} = a_2^2 - 3{a_1}{a_3}$, ${B_2} = {a_1}{a_2} - 9{a_0}{a_3}$, $D = (2{B_1}{a_2} - 3{a_3}{B_2})/(2B_1^{3/2})$ and $ \theta  = \arccos D/3$.

Step II: Establish the algebraic interrelation between the parameters of $\eta ,{w_2},{C_{{g_0}}}$ and the asymptomatic quadratic-growth-coefficients of $\alpha _2^0,\alpha _2^1,\alpha _1^1$ with the solutions obtained in step I to give finally
\begin{equation}\small
\left\{\begin{array}{l}
 {w_2}(\eta ,\alpha _2^0,\alpha _2^1) = ({T^2} + 6\eta \alpha _2^1T)/[36\eta {(\alpha _2^1)^2} + 6\eta \alpha _2^1T] \\
\alpha _1^1(\eta ,\alpha _2^0,\alpha _2^1) = \alpha _2^1 \chi _{{g_0}}^1\\
 {C_{{g_0}}}(\eta ,\alpha _2^0,\alpha _2^1) = 3{w_2}\eta /[2\alpha _2^0(1 + \eta )] - 1/2 \\
 \eta ({w_2},\alpha _2^0,\alpha _2^1)=[-{b_2}-({b_2^2-4{b_1}{b_3}})^{1/2} ]/(2{b_1}) \\
 \eta (\alpha _1^1,\alpha _2^0,\alpha _2^1) =[-{d_2}-({d_2^2 - 4{d_1}{d_3}})^{1/2} ]/(2{d_1})
 \end{array} \right.,
\label{eq6}
\end{equation}
where $\chi _{{g_0}}^1=[(1 - {w_2})\eta +({{{(1 - {w_2})}^2}{\eta ^2} + 4{w_2}\eta })^{1/2} ]/$ $(2{w_2}\eta )$, $T = {e_1} + {e_3}\eta$, ${b_1} = e_3^2 + {e_2}{e_3}(1 - {w_2})$, ${b_2} = 2{e_1}{e_3} + {e_1}{e_2}(1 - {w_2}) - e_2^2{w_2}$ ,${b_3} = e_1^2$, ${d_1} = \mu (\mu+ 1)e_3^2 + {e_2}{e_3}{\mu ^2}$,  ${d_2} = 2{e_1}{e_3}\mu (\mu  + 1) + {e_1}{e_2}{\mu ^2} - {e_2}{e_3} - e_2^2\mu$, ${d_3} = e_1^2\mu (\mu  + 1) - e_2^2 - {e_1}{e_2}$   , ${e_1} = \alpha _2^0(3 + 2\alpha _2^1),{e_2} = 6\alpha _2^1,{e_3} = {e_1} - {e_2}$ and $\mu  = \alpha _1^1/\alpha _2^1$.

Step III: Determine the values of the parameters. First, the range of $\eta\in (\eta_{min}$, $\eta_{max})$ is determined by combining the physical constraints of ${w_2}<1/2$ (see Fig.{\ref{fig1}}), $\alpha _1^1<1/2$ (restricted by free fall\cite{Dimonte2000}) and the asymptotic requirements of $\alpha _2^0 = 0.06$ (see experiments \cite{Youngs1989,Dimonte2000exp,Schneider1998}), $\alpha _2^1 = 0.05$ (see theories \cite{Alon1994,Alon1995,Glimm1990}). In fact, the first two expressions of Eq. ({\ref{eq6}}) show that ${w_2}$ ($\alpha _1^1$) is a decreasing (increasing) function of $\eta$ nearby $\eta=1$, thus giving ${\eta _{\min }}({w_2} = 1/2$, $\alpha _2^0$, $\alpha _2^1) = 0.56$ and ${\eta _{\max }}(\alpha _1^1 = 1/2$, $\alpha _2^0$, $\alpha _2^1) = 1.37$. Second, by using the first and the third expressions of Eq.(\ref{eq6}), one can calculate ${w_2}$ and ${C_{{g_0}}}$ by assigning specific $\eta$ to give specially ${w_2}(\eta  = 1$, $\alpha _2^0$, $\alpha _2^1) = 0.24$ and ${C_{{g_0}}}(\eta  = 1$, $\alpha _2^0$, $\alpha _2^1) = 2.5$.

Step IV: Utilize the obtained ${C_{{g_0}}}$ to determine the drag coefficient $C$ for the variable $g(t)$ problem, for which a time-dependent value of $C[g(t)]$ is expected to be obtained. The main logic can be summarized as follows. In physics, drag is proportional to the surface area of the bubble structure \cite{Alon1995,Cheng2002} (denoted as $S$) and is highly dependent on the direction of mixing. For a case in which  ${v_2} \ge 0$, the drag is dominated by chunk mixing near the local mixing front, and the experiments further imply a negative correlation between $S$ and $dg/dt$ \cite{Dimonte1996,Dimonte2000exp}, leading to $C = {C_{{g_0}}}$, $C\approx{C_{{g_0}}}$  , $C< {C_{{g_0}}}$, $C >{C_{{g_0}}}$ for problems driven by constant, oscillating, increasing, and decreasing $g(t)$, respectively. In contrast, for cases in which ${v_2} < 0$, the drag is dominated by atomic mixing \cite{Livescu2011} across the entire BMZ ($S\gg {S_{{v_2} \ge 0}}$), leading to $C \gg{C_{{g_0}}}$.

With the parameters determined above, our theory is systematically validated for general RT mixing, as shown in Fig. 2-4. Fig.2 shows the validation for an RT problem with constant $g$ and at all $R$, where our predictions are in good agreement with the results of the experiments \cite{Dimonte2000exp}. In our predictions, the different values of $\eta$ are used to reveal the dependence of the evolutions on the symmetry (breaking) of the density fields, to reproduce the many observed pairs of  $\alpha _i^A$ at the same $R$, and to explain the distinct evolutions in two experiments with the same configurations \cite{Youngs2013}. Fig.2 further indicates that: (1) Except for the two over-measured points, almost every observed $\alpha _1^A$ is within the region bounded by curves with the maximum and minimum $\eta$; (2) Except for very few experiments, the density fields in BMZ and SMZ are symmetrical approximately in most cases since the majority of the observed $\alpha _i^A$ values lie on curves for which $\eta  \approx 1$; (3)$\alpha _{1(2)}^A$, or equivalently $h_{1(2)}(t)$, is closely (slightly) dependent on $\eta$, consistent with the results of the experiments; (4)$\alpha _1^A/\alpha _2^A$ is closely dependent on $\eta$, explaining the distinct difference between the empirical formula of $\alpha _1^A/\alpha _2^A \sim {R^{0.33}}$ observed in linear electric motor experiments \cite{Dimonte2000exp} and that of $\alpha _1^A/\alpha _2^A \sim {R^{0.23}}$ in rocket-rig experiments \cite{Youngs1989,Youngs2013} for the first time.

\begin{figure}[t]
\centering
\includegraphics[width=8.5cm,trim=0.15cm 0.00cm 2.00cm 1.00cm,clip]{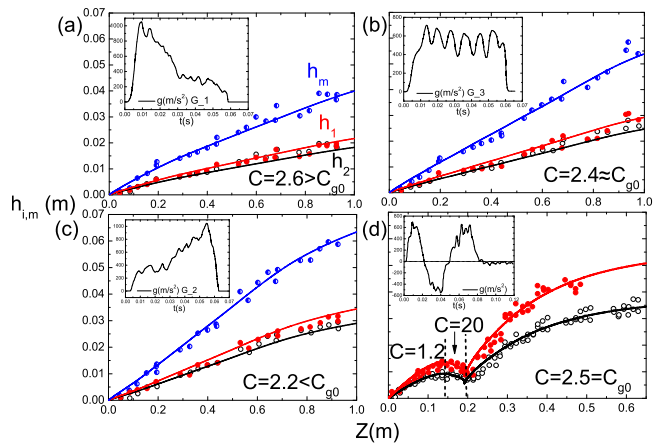}
\caption{ \label{fig3} (color online). The comparison of $h_1(t)$ (red), $h_2(t)$ (black), and ${h_m}(t) \equiv {h_1}(t) + {h_2}(t)$ (blue) between theoretical predictions (lines) and experiments (symbols \cite{Dimonte2000exp,Dimonte2007}) for problems driven by (a) increasing,(b) oscillating,(c) decreasing  and (d) complex  $g(t)$ (shown in inset). $R=1.57$ in cases (a)-(c), and $R=2.83$ in case (d). $Z$ is defined as $Z\equiv \int\int g(t')dt'dt$. The lines are obtained by integrating Eq.(\ref{eq4}) with $\eta  = 1$ and the suggested $C$ (see text and figures). Specifically, for case (d), according to the sign of $v_2$, the piecewise constant $C$ are used.}
\end{figure}

Figs. 3-4 show a validation for problems with diverse $g(t)$ and at all $R$ ($v_{i0}=0,h_{i0}=10^{-6}m$ \cite{Dimonte2000}). For variable $g(t)$ problems, although $C[g(t)]$ has not yet been formulated, we can still predict ${h_i}(t)$ with a reasonable approximation of $C = const.$, either entirely or piecewise. By means of this approximation, our theory is in good agreement with the results of experiments, and much better than theory given in Ref.\cite{Dimonte2000} (illustrated by the example in the inset of Fig. 4). To the best of our knowledge, this is the first time that a theory has successfully reproduced the results of all the experiments with the same parameters determined definitely.

\begin{figure}[t]
\centering
\includegraphics[width=8.5cm,trim=0.00cm 0.00cm 0.00cm 0.50cm,clip]{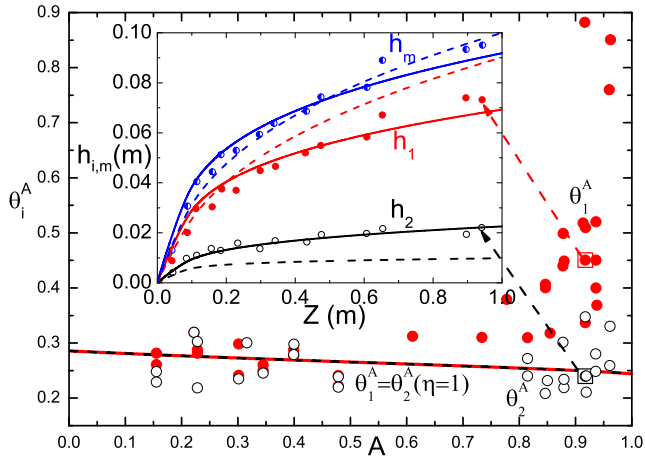}
\caption{ \label{fig4} (color online). The comparison of power-index  $\theta _1^A$(red) and $\theta _2^A$(black) between theoretical predictions (lines) and experiments (symbols \cite{Dimonte2000exp}) for problems with impulsive $g(t)$ and at all $R$. The lines of $\theta _1^A = \theta _2^A(\eta  = 1)$ show the exact solution for the special initial condition (see text). For general cases, Eq.(\ref{eq4}) successfully reproduces all the impulsive experiments conducted by Dimonte \cite{Dimonte2000exp} but only one example ($R = 23.4$ ,F\_68,G15 \cite{Dimonte2000exp}) is shown in the inset. The inset compares the results of experiment \cite{Dimonte2000exp} with predictions by Eq.(4) (solid lines) and by Dimonte's model (dashed lines \cite{Dimonte2000}). The current prediction is conducted by integrating Eq.(\ref{eq4}) with experimentally measured $g(t)$\cite{Dimonte2000exp} and parameters of $\eta$, $w_2(\eta)$ and $C_{g_0}(\eta)$ , where a constant $C_{g_0}(\eta)$ is adopted to neglect the variation of $C[g(t)]$ in stage I. In a very few experiments (especially those with a large $R$), probably affected by distinct initial perturbations, the $\eta$ needed to be adjusted slightly around 1, as in the inset with $\eta  = 1.1$.}
\end{figure}

As a special example of variable $g(t)$ problem, it is necessary to validate and analysis the current theory for problem with impulsive $g(t) = U\delta (t)$ (Richmyer-Meshkov mixing \cite{Dimonte2000,Dimonte2000exp}). To this end, we have divided the entire $g(t)$ into stage I with impulsive acceleration and stage II with zero acceleration \cite{Dimonte2000exp}. Given the extremely short duration of stage I, we only investigated stage II and have denoted the quantities at the end of stage I with subscript $\delta$. For stage II with initial values of ${v_{i\delta }}$  and ${h_{i\delta }}$, Eq.(\ref{eq4}) approximately has a power-law solution \cite{Dimonte2000} of
\begin{eqnarray}
{h_i} = {h_{i\delta }}(\frac{{t}}{{{t_{i\delta }}\theta _i^A}}+ 1)^{\theta _i^A},{t_{i\delta }} = \frac{{h_{i\delta }}}{{v_{i\delta }}},\theta _i^A = \theta _{i\delta }^A[1 + \varepsilon (t)],
\label{eq7}
\end{eqnarray}
where $\varepsilon (t)$ is generally a time-dependent small quantity \cite{Dimonte2000}. The solution enables us to understand some long-standing questions. First, for the special initial condition of ${h_{1\delta }}/{h_{2\delta }} = {v_{1\delta }}/{v_{2\delta }} = \alpha _1^A/\alpha _2^A$, one can verify that $\varepsilon (t)$ equals zero exactly, with the corresponding power-index of $\theta _1^A = \theta _2^A = \theta _{2\delta }^A$, $\theta _{2\delta }^A = 1/[1 + C\phi (R,\eta ,{\chi _\delta })]$ (see Ref.\cite{Dimonte2000} for more information). This exact solution can explain the observed  $\theta _1^A \approx \theta _2^A$ for $A<0.8$ (see Fig.4) as follows: for a problem with a small/moderate $R$ and a positive $g(t)$, previous studies \cite{Dimonte1996,Dimonte2000exp} suggested an empirical formula of ${h_i} \approx \alpha _i^AA{[\int {\sqrt {g(t)} dt} ]^2}$ which, when applied to stage I, gives the special initial condition, leading to $\theta _1^A \approx \theta _2^A$. Second, for general cases, if $t \gg{t_{i\delta }}$ and we neglect the high-order modification by $\varepsilon (t)$ \cite{Dimonte2000}, we can obtain $\theta _2^A \approx \theta _{2\delta }^A(\phi)$ explicitly by following the procedures given in Ref.\cite{Dimonte2000} and $\theta _1^A \approx f({\chi _\delta },{t_{i\delta }},\eta )$ implicitly by substituting the power-law solution into the first equality of Eq.(\ref{eq4}). The former reveals that $\theta _2^A$, or equivalently ${h_2}(t)$, is nearly independent of the initial conditions, and determined dominantly by $\eta$ because $\phi (R,\eta ,{\chi _\delta })$ is only slightly dependent on ${\chi _\delta }$ (this can be verified with the expression of $\phi$). In contrast, the latter implies that $\theta _1^A$, or equivalently $h_1(t)$, is sensitive to $\eta$ and the initial conditions, as verified by numerical integration. These conclusions are consistent with the experimental \cite{Dimonte2000exp} and theoretical \cite{Zhang1998} results. Finally, in the same way as the inference in the problem with constant $g$, the good agreement of $\theta _i^A$  with $\eta  = 1$ in both the experiments and solutions (see the lines in Fig.4) implies that the density fields in BMZ and SMZ are symmetrical approximately in most experiments, too.

A discussion is in order. Our theory was validated systematically in terms of reproducing the results of all the available experiments, but only those results obtained for systems with immiscible inviscid fluids \cite{Youngs1989,Alon1994,Alon1995,Schneider1998,Dimonte2000exp} and natural perturbations \cite{Youngs1989,Schneider1998,Dimonte2000exp} are presented in this letter. As is well known, however,$\alpha _2^A$ is highly dependent on the fluid properties (such as viscosity, miscibility)\cite{Dimonte2000,Dimonte2004} and initial perturbations \cite{Dimonte2004pre,Youngs2013,Ramaprabhu2005}. Therefore, for other systems with notably different media and/or perturbations, slightly different values of parameters $\alpha _2^0$ and/or $\alpha _2^1$  may be used. Nevertheless, the good agreements substantially confirm that the evolutions of $h_i(t)$ are governed essentially by conservation principle. As for the symmetry principle, the $\eta$ in the current theory is introduced as a free parameter to reproduce, explain, and reveal the different evolutions in experiments using the same $R$ and $g(t)$, and a strong (weak) dependence of ${h_{1(2)}}(t)$  on $\eta$ is found. Although $\eta$ may depend on many factors, we infer that the initial perturbations promise to be the most important factors \cite{Burrows2000}. Finally, Fig. 2-4 show that the density fields in BMZ and SMZ are of symmetry breaking in most experiments where $\eta$ does not equal unity exactly, probably a result of natural perturbation. Therefore, it is reasonable to infer that the symmetry breaking may be universal in nature, and current work further demonstrates the possibility of applying the concept to understand the intractable turbulence problem. For other problems, the concept may work, too.

This work was supported in part by CAEP under Grant Number 2012A020210, and from NSFC under Grant Numbers 11502029, 11572052, 11472059, 11171037, 11372330, 11472278 and 91441103.

%\nocite{*}

%\bibliography{M_invariant_MVP_of_CTBLs}% Produces the bibliography via BibTeX.

\end{document}